\documentclass[galaxies,article,accept,moreauthors,pdftex,10pt,a4paper]{mdpi}

\firstpage{1}
\articlenumber{34}
\doinum{10.3390/galaxies5030034}
\pubvolume{5}
\pubyear{2017}
\copyrightyear{2017}
\externaleditor{Academic Editors: Duncan A. Forbes and Ericson D. Lopez}
\history{Received: 1 July 2017; Accepted: 1 August 2017; Published: 4 August 2017}

\usepackage{amssymb}
\usepackage{booktabs}
\usepackage{multirow}
\usepackage{soul}
\usepackage{microtype}
\graphicspath{ {Figures/}}

\newcommand{\apj}{ApJ}			

\newcommand{\nat}{Nature}		

 \theoremstyle{mdpi}
 \newcounter{thm}
 \newcounter{ex}
 \newcounter{re}
\Title{Galaxies with Shells in the Illustris Simulation: Metallicity Signatures}

\Author{{Ana-Roxana Pop} $^{1,}$*\orcidA{},
Annalisa Pillepich $^{1,2}$, Nicola C. Amorisco $^{1,3}$ and Lars Hernquist $^{1}$}
\AuthorNames{Ana-Roxana Pop, Annalisa Pillepich, Nicola C. Amorisco, and Lars Hernquist}

\address{%
$^{1}$ \quad Harvard-Smithsonian Center for Astrophysics, 60 Garden St., Cambridge, MA 02138, USA; pillepich@mpia-hd.mpg.de (A.P.); nicola.amorisco@cfa.harvard.edu (N.C.A.); lhernquist@cfa.harvard.edu~(L.H.)\\
$^{2}$ \quad Max-Planck-Institut f{\"u}r Astronomie, K{\"o}nigstuhl 17, 69117 Heidelberg, Germany\\
$^{3}$ \quad Max Planck Institute for Astrophysics, Karl-Schwarzschild-Strasse 1, D-85740 Garching, Germany}

\corres{Correspondence: ana-roxana.pop@cfa.harvard.edu}

\abstract{Stellar shells are low surface brightness arcs of overdense stellar regions, extending to large galactocentric distances.
In a companion study, we identified 39 shell galaxies in a sample of 220~massive ellipticals \mbox{($\mathrm{M}_{\mathrm{200crit}}>6\times10^{12}\,\mathrm{M}_\odot$)} from the Illustris cosmological simulation.
We used stellar history catalogs to trace the history of each individual star particle inside the shell substructures, and we found that shells in high-mass galaxies form through mergers with massive satellites (stellar mass ratios $\mu_{\mathrm{stars}}\gtrsim1:10$).
Using the same sample of shell galaxies, the current study extends the stellar history catalogs in order to investigate the metallicity of stellar shells around massive galaxies. Our results indicate that outer shells are often times more metal-rich than the surrounding stellar material in a galaxy's halo. For a galaxy with two different satellites forming $z=0$ shells, we find a significant difference in the metallicity of the shells produced by each progenitor.
We also find that shell galaxies have higher mass-weighted logarithmic metallicities ([Z/H]) at $2$--$4\,\mathrm{R}_{\mathrm{eff}}$ compared to galaxies without shells.
Our results indicate that observations comparing the metallicities of stars in tidal features, such as shells, to the average metallicities in the stellar halo can provide information about the assembly histories of galaxies.}


\begin{document}

\section{Introduction}
Stellar halos are diffuse, low surface brightness regions
that contain unique morphological features, such as tidal streams, stellar shells, rings, or plumes (e.g., \cite{McConnachieetal2009, MartinezDelgadoetal2010}).
Shells are a particular type of tidal debris that appear as interleaved caustics with large opening angles, which have been identified in numerous deep surveys probing the faint outskirts of galaxies \cite{Arp1966, Malin&Carter1983, Schweizer1983, Schweizer&Ford1985, Taletal2009,   Atkinsonetal2013, Krajnovicetal2011, Duc2016}.	
Interest in the detailed study of stellar substructures such as shells has been stimulated by previous work indicating that tidal features are powerful tracers of galaxies' assembly histories (see e.g.,   \cite{Arnaboldietal2012, Martinez-Delgadoetal2012, Romanowskyetal2012, Fosteretal2014, Longobardietal2015, Amoriscoetal2015}).
Due to the long dynamical timescales in the outskirts of galaxies,
information about the accretion histories of galaxies is well preserved at the present epoch. Previous studies indicate that stellar shells are very long-lived structures, with average lifetimes of $\sim$2--3 Gyr according to a recent paper \cite{Popetal2017}.

Several theories have been proposed for the formation of shells, ranging from models invoking star formation in shocked galactic winds \cite{Fabianetal1980, Bertschinger1985, Williams&Christiansen1985, Loewensteinetal1987}
to tidal interaction theories in which the shells are density waves induced by a passing galaxy \citep{Thomson&Wright1990, Thomson1991}.
Nevertheless, extensive numerical and observational studies support the merger scenario, in which shells are composed of stars stripped from accreted satellites (e.g., \cite{Quinn1984, Dupraz&Combes1986, Hernquist&Quinn1988,Hernquist&Quinn1989, Sanderson&Bertschinger2010, Sanderson&Helmi2013}).
The increasing richness of stellar halo observations will give us a unique look into the past merger histories of galaxies, yet the interpretation of the data requires detailed model predictions for the formation of stellar substructures. Most of the previous studies on shell galaxies have focused on idealized mergers involving small satellites on radial orbits, (e.g., \citep{Hernquist&Quinn1987b, Hernquist&Quinn1988, Hernquist&Quinn1989, Dupraz&Combes1986, Seguin&Dupraz1996, Sanderson&Bertschinger2010, Bartoskovaetal2011, Ebrova2012}), yet
major mergers have also been shown to produce shells, (e.g., \cite{Hernquist&Spergel1992, GonzalezGarcia&Balcells2005,GonzalezGarcia&vanAlbada2005a, GonzalezGarcia&vanAlbada2005b}).
In a companion paper, \mbox{Pop et al. (\cite{Popetal2017}, hereafter P17)} found that shells in high-mass galaxies are often the result of mergers with
massive satellites (relative to the mass of the host galaxy).
Since the efficiency of dynamical friction is higher for major mergers compared to minor mergers, massive satellites can form shells for a wide range of impact parameters; small satellites, however, need almost perfectly radial orbits in order to produce stellar shells.
As a result, in P17, we show that most satellites producing shells in massive ellipticals have stellar mass ratios $\mu_{\mathrm{stars}}\gtrsim1:10\,$.

Previous studies suggest that metallicity gradients can be used to
characterize the relative importance of major and minor mergers in the accretion history of a galaxy (e.g., \cite{Hirschmann2015,Kobayashi2004}). During~the initial formation phase of massive galaxies, in-situ star formation will leave a steep metallicity gradient, with stars on the outskirts of galaxies forming out of lower density gas and metals being further removed from the outer regions of the stellar halo by stellar winds \cite{Carlberg1984,Kobayashi2004}. Overall, subsequent accretion events ($z \lesssim 1$)
have the net effect of flattening the metallicity profiles (see e.g., \cite{BekkiShioya1999, DiMatteo2009, Font2011}).
In a previous study of Illustris galaxies, Cook et al. \cite{Cooketal2016} found that galaxies with higher fractions of accreted material tend to have less steep gradients than galaxies with a lower fraction of ex-situ stars.
Based on these results, it has been suggested that metallicity gradients could provide additional information about the mass ratios of mergers that produce shells. Several groups have found that shell galaxies tend to have lower central Mg$_2$ values than non-shell galaxies (e.g., \cite{Longhetti2000,Carlstenetal2016}), with galaxies with shells lying below the Mg$_2$--$\sigma$ relation from \cite{Bender1993}.

\textls[-5]{Studies of shell galaxies using deep imaging face several challenges, such as the
extremely low surface brightness of shells ($\sim$$28 \; \mathrm{mag} \; \mathrm{arcsec}^{-2}$ or fainter in the V band, e.g., \cite{Johnston2008, Atkinsonetal2013}),
the varied and sometimes irregular shell morphologies,
and the fact that outer shells extend far out in the outskirts of the stellar halo, at galactocentric distances $\sim$$100\, \mathrm{kpc}$.
For sufficiently nearby elliptical galaxies ($\leq$$20\, \mathrm{Mpc}$), direct stellar photometry has successfully provided strong constraints on the age and metallicity of stars extending to large galactocentric distances in shell galaxies (e.g., \cite{Rejkubaetal2005, Rejkubaetal2011, Rejkubaetal2014, Mihosetal2013}).
In this paper, we investigate the metallicity of stellar shells in the Illustris simulation, studying how different metallicity signatures can set constraints on the number and mass of shell-forming progenitors.
As~metallicity measurements are reaching larger radii through slit spectroscopy (e.g.,~\cite{San2007, Foster2009, Spolaor2010, Coccato2010, Coccato2011}), integral-field spectroscopy \linebreak (e.g.,~\cite{Kuntschner2010, Weijmans2009, Greene2015, GonzalezDelgado2015, Wilkinson2015}), and~deep photometric studies (e.g.,~\cite{Harrisetal2007, Mihosetal2013, Rejkubaetal2014, Lee&Jang2016}),
as well as probing increasingly larger samples of galaxies \cite{Spolaor2010, Pastorello2014}, the study of stellar metallicities of tidal features will provide previously inaccessible information about the galaxies' accretion histories.}

\section{Simulations and Methods}

\subsection{Illustris}

This project uses the Illustris simulation \citep{Vogelsbergeretal2014a, Vogelsbergeretal2014b, Geneletal2014}, a cosmological hydrodynamical simulation run using {\small AREPO} \citep{Springel2010}---a moving-mesh code based on a quasi-Lagrangian finite volume method. We are working with the highest resolution run ({Illustris-1}), which corresponds to a simulation box with a periodic volume of $(106.5\;\mathrm{ Mpc})^3$ and mass resolution of $\mathrm{m}_{DM} = 6.26 \times 10^6\, \mathrm{M}_{\odot}$.
The Illustris simulations include a broad range of astrophysical processes (e.g., feedback from supermassive black holes, supernovae, and AGNs) in order to self-consistently reproduce
galaxy formation and \mbox{evolution \citep{Vogelsbergeretal2013, Torreyetal2014}}.

\subsection{The Galaxy Sample}

Similar to P17, we start with an initial sample consisting of the 220 most massive central galaxies in Illustris at redshift $z=0$.
This corresponds to a virial mass cut $\mathrm{M}_{\text{200crit}}>6 \times 10^{12}\, \mathrm{M}_{\odot}$, where $\mathrm{M}_{\text{200crit}}$ is defined as the total mass inside a radius enclosing a sphere with mean density $200$ times greater than the critical density of the Universe.
In Illustris, central galaxies are the most massive galaxies in a given halo/group, and
satellites inside each halo are identified using the {\small SUBFIND} algorithm \citep{Davisetal1985, Springeletal2001, Dolagetal2009}.

\subsection{Stellar History Catalogs and Shell Galaxies Identification}
\label{subsec:historyCat}
{The present study uses the Illustris shell galaxies previously identified in P17 using a two-step~approach:}
\begin{itemize}[leftmargin=*,labelsep=5.8mm]
\item	\textit{Step 1}: Galaxies with stellar shells are visually identified using stellar surface density maps. Each~galaxy in the sample is studied using all three projections ($x$--$y$, $y$--$z$, $z$--$x$) and two different contrast levels, and each image stamp received a score ranging from $0$--$2$ from three different members of our team. A score of 2 indicates a galaxy with two or more well-defined shells, a~score of 1 corresponds to galaxies with one or two shell-like structures, and a score of 0 indicates no shell detection.
We order candidate shell galaxies based on the total scores given to all their corresponding image stamps. As discussed in P17, most shell galaxies exhibit shell-like structures in at least 2/3 projections, and we use the second identification step to verify the presence of shells in each of the candidate galaxies.

\item	\textit{Step 2}: We develop stellar history catalogs
in order to identify the shell-forming progenitors and separate stars in shells from all other stars in the galaxy.
In these catalogs, we trace the birth, trajectory and progenitors of all
star particles inside $z=0$ halos, saving information about each individual star particle at three key moments during its life: formation, accretion (when the parent satellite enters the virial radius of the host) and stripping time (when the star becomes gravitationally bound to the new host).
In P17, we show that shells in Illustris are composed of ex-situ stars, which is in agreement with merger models predicting that shell-like structures correspond to overdensities of stripped stars accumulating at the apocenters of their orbits (see~analytical and numerical studies by, e.g., \cite{Quinn1984, Dupraz&Combes1986, Hernquist&Quinn1988, Hernquist&Quinn1989, Hernquist&Spergel1992, Sanderson&Bertschinger2010, Sanderson&Helmi2013, Amorisco2015, Hendel&Johnston2015}).
By tracing the configuration and phase space ($v_r$ vs. $r$) distribution of star particles
with a common parent satellite,
we obtain a systematic sample of all the satellites that are responsible for $z=0$ stellar shells.
\end{itemize}

Out of 220 massive ellipticals in Illustris (with $\mathrm{M}_{\text{200crit}}>6 \times 10^{12}\, \mathrm{M}_{\odot}$), Pop et al. \cite{Popetal2017} find that~39~galaxies exhibit stellar shells visible at $z=0$, corresponding to a fraction of $18\% \pm 3\%$ of all galaxies in the sample.\footnote{We provide Poisson errors for the fractions of galaxies with shells in Illustris.

However, the actual uncertainties in the $f_{\mathrm{shells}}$ could be higher due to environment (rich groups and clusters vs. field galaxies), mass distribution of galaxies in the sample, redshift evolution $f_{\mathrm{shells}}(z)$, surface brightness limits, and projection effects. For more details about the impact of these effects, see the discussion in Pop et al. \cite{Popetal2017}.} Moreover, the fraction of galaxies with shells increases with increasing mass cut, with as many as $34\% \pm 11\%$ of galaxies with $\mathrm{M}_{\text{200crit}}>3 \times 10^{13}\, \mathrm{M}_{\odot}$ having shells. Despite~significant uncertainties in the number of observed shell galaxies, the incidence of galaxies with shells in Illustris is in overall agreement with current observational limits (e.g., \cite{Malin&Carter1983, Schweizer1983, Schweizer&Ford1985, Taletal2009, Atkinsonetal2013}).

\begin{figure}[H]
\centering
\includegraphics[width=\textwidth]{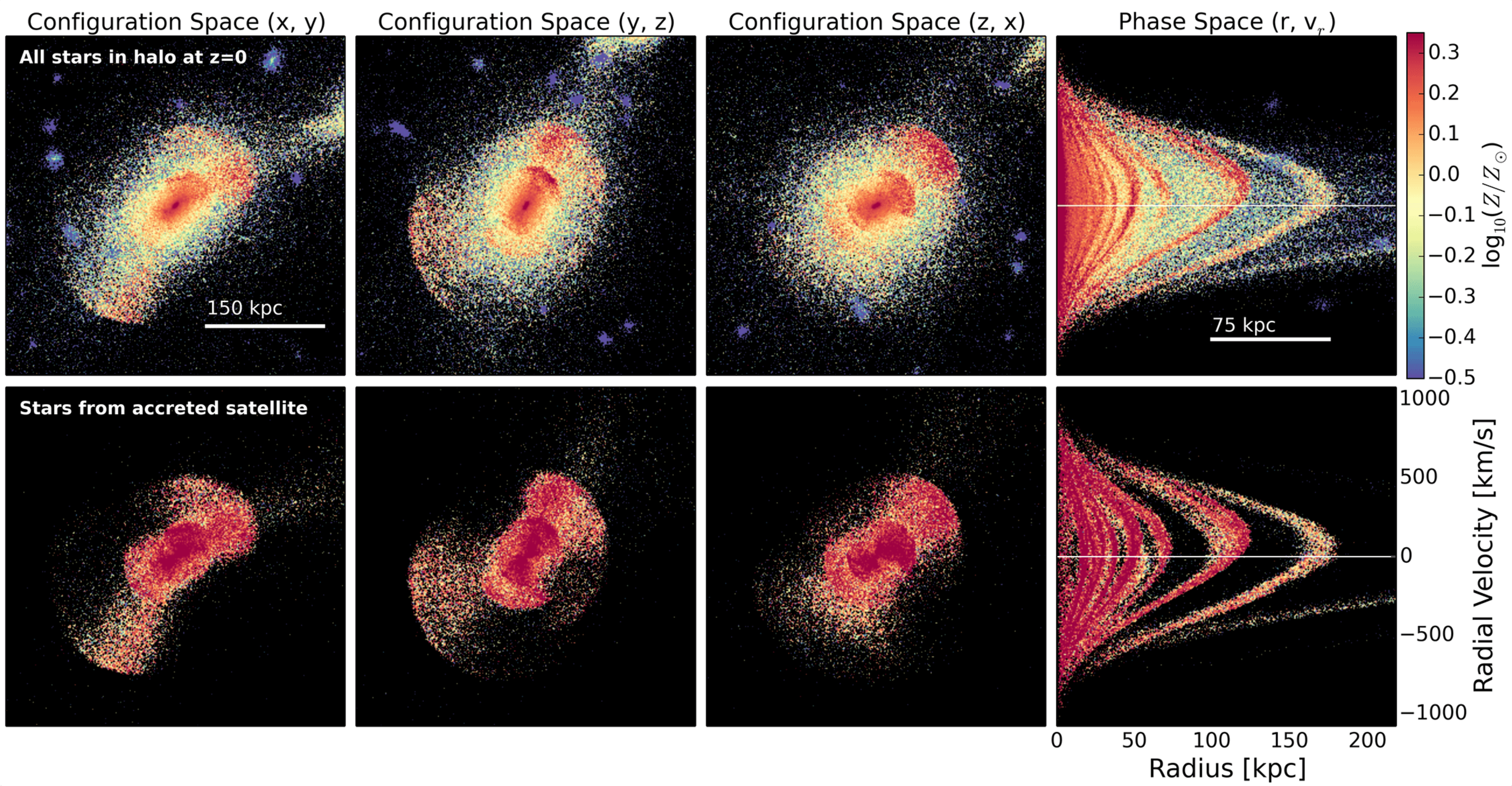}
\caption{\textls[-15]{Stellar maps in configuration space (first three columns) and phase space, i.e., radial velocity vs. radial distance (last column).
The color of each bin of area $\mathrm{A}_{\mathrm{bin}} = (1.5\, \mathrm{kpc})^2$ in the 2D histograms above corresponds to the median total metallicity (Z) of all star particles therein.
\textbf{Top row}:
entire halo of a redshift $z=0$ shell galaxy with total stellar mass $\mathrm{M}_\mathrm{stars} = 6.3 \times 10^{11}\, \mathrm{M}_\odot$, virial mass $\mathrm{M}_\mathrm{200crit} = 1.4 \times 10^{13}\, \mathrm{M}_\odot$, and effective radius $\mathrm{R}_{\mathrm{eff}} = 16.3\, \mathrm{kpc}$ (where $\mathrm{R}_{\mathrm{eff}}$ is defined as the radius that contains half of the total light in a galaxy).
\textbf{Bottom row}: redshift $z=0$ stars that have been accreted from the same common progenitor, which is responsible for the shell structures in the top row.
We find that stellar shells in this galaxy have higher total metallicities than stars in the outer regions of the stellar halo.
Moreover, we find a gradient in the average metallicity of individual shells, with outer shells having lower average Z that inner shells.}}
\label{fig:metallicityShellsExample1}
\end{figure}

In this paper, we take advantage of the versatility of stellar history catalogs, which can be extended to study a wide range of properties for stars inside tidal features like shells, streams or plumes (for~more details about the construction of the catalogs, see P17). In particular, the current study focuses on the metallicity signatures of several shell galaxies from the Illustris simulation.
We use stellar surface density maps such as those presented in Figure~\ref{fig:metallicityShellsExample1}, in which we weigh each star particle in the $z=0$ halo by its total metallicity (Z), normalized to that of the Sun ($Z_\odot$).
Each bin in the 2D histograms in~Figure~\ref{fig:metallicityShellsExample1} is colored based on the median total metallicity of all star particles therein.
In~Illustris, the metallicity of each individual star particle is based on the metallicity of the gas cell at the star's time of birth (i.e., when the gas cell is converted into a star particle).

\begin{figure}[H]
\centering
\includegraphics[width=\textwidth]{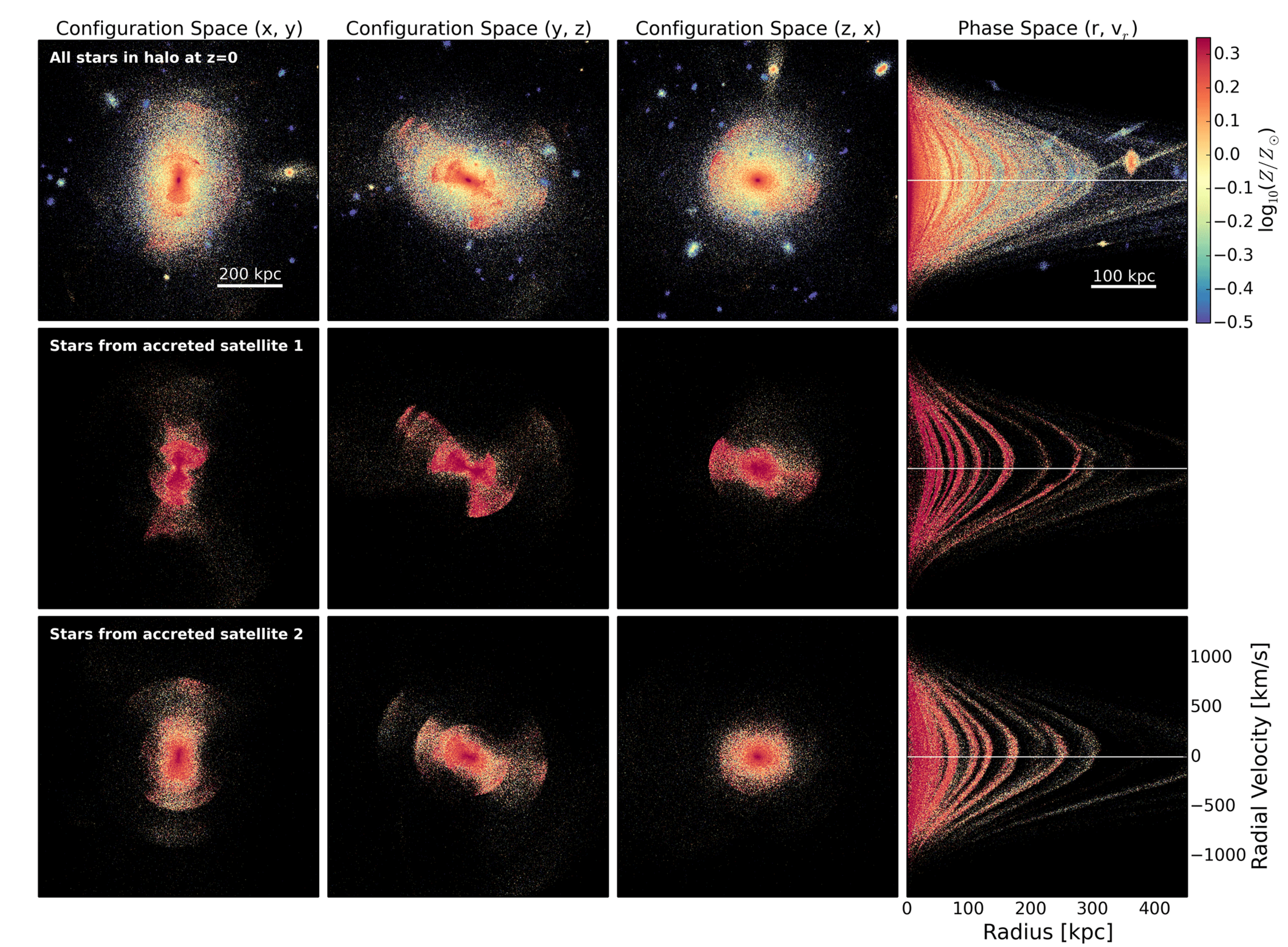}
\caption{Stellar maps in configuration space (first three columns) and phase space, i.e., radial velocity vs. radial distance (last column).
The color of each bin of area $\mathrm{A}_{\mathrm{bin}} = (1.5\, \mathrm{kpc})^2$ in the 2D histograms above corresponds to the median total metallicity (Z) of all star particles therein.
\textbf{Top row}: entire halo of a $z=0$ shell galaxy with total stellar mass $\mathrm{M}_\mathrm{stars} = 1.3 \times 10^{12} \,\mathrm{M}_\odot$, virial mass $\mathrm{M}_\mathrm{200crit} = 4.6 \times 10^{13}\, \mathrm{M}_\odot$, and $\mathrm{R}_{\mathrm{eff}} = 33.5 \,\mathrm{kpc}$. This galaxy had two different shell-forming progenitors responsible for the shells visible at $z=0$.
\textbf{Middle and bottom rows}: redshift $z=0$ stars accreted from each of the two shell-forming satellites with stellar mass ratios $\mu_{\mathrm{stars}} = \mathrm{M}_{\mathrm{satellite}}^{\mathrm{stars}} / \mathrm{M}_{\mathrm{host}}^{\mathrm{stars}} = 0.15$ and $0.24$, measured at accretion time (i.e., when the satellite comes inside the virial radius of the host).
Stellar shells in this galaxy have higher average metallicity than the other stars in the outskirts of the stellar halo. Moreover, the shells produced by the first satellite are on average more metal-rich than those produced by the second satellite.
Similar to Figure~\ref{fig:metallicityShellsExample1}, we find that the outer shells have lower total metallicities that inner shells.}
\label{fig:metallicityShellsExample2}
\end{figure}

In configuration space, we identify shells as interleaved stellar overdensities, often observed on both sides of the galaxy center and extending to very large galactocentric distances, in the low surface brightness regions of the stellar halo.
Stellar shells also have a specific signature in phase space, i.e.,~in the space of galactocentric distance vs. radial velocity ($r$--$v_r$).
Stars spend longer times near the apocenters of their orbits, where their radial velocities are close to zero, thus creating phase wraps peaking at $v_r \simeq 0$, as exemplified in the right column of Figure~\ref{fig:metallicityShellsExample1}.
Using stellar history catalogs, we~identify satellites responsible for $z=0$ shells, and we investigate the metallicity distribution of stars inside these shells (see bottom rows of Figures~\ref{fig:metallicityShellsExample1}
and \ref{fig:metallicityShellsExample2}).

\section{Results and Implications}

We begin by investigating the distribution of stellar metallicities for two shell galaxies, one~for which the $z=0$ shells are composed of stars accreted from a single satellite, while the shells in the second galaxy are the result of mergers with two different satellites.
In Figure~\ref{fig:metallicityShellsExample1}, we study the configuration and phase space distribution of all star particles inside the halo of a galaxy with total stellar mass $\mathrm{M}_\mathrm{stars}(z=0) = 6.3 \times 10^{11}\, \mathrm{M}_\odot$ and virial mass $\mathrm{M}_\mathrm{200crit}(z=0) = 1.4 \times 10^{13}\, \mathrm{M}_\odot$.
The~shell-forming satellite depicted in the bottom row corresponds to a close to $1:1$ merger (in~total mass) and a stellar mass ratio of $\mu_{\mathrm{stars}} = \mathrm{M}_{\mathrm{satellite}}^{\mathrm{stars}} / \mathrm{M}_{\mathrm{host}}^{\mathrm{stars}} = 0.22\,$. The~satellite was accreted $\sim$4 Gyr ago on a low angular momentum orbit, with a radial velocity ratio\footnote{The radial velocity ratio ($v_r/|v|$) is defined as the radial component
of the relative velocity of the satellite ($v_r$), normalized to the modulus of the relative velocity ($v$).} at accretion time of $v_r/|v| = -0.97$.
This shell-forming satellite has properties that agree with the overall trends for shell formation found in P17:
most of the shell galaxies in the current sample form through mergers
with stellar mass ratios  $\mu_{\mathrm{stars}} \gtrsim 1:10$,
and the shell-forming satellites are accreted on low angular momentum orbits, at~intermediate accretion times ($\sim$4--8 Gyr ago).
We determine the color of each bin in the stellar 2D histograms in Figure~\ref{fig:metallicityShellsExample1} by computing the median of the total metallicity (Z) for all star particles in that bin.
We find metal-rich shells visible in all three projections and that extend to large galactocentric distances (>50 kpc).
While in the top panels of Figure~\ref{fig:metallicityShellsExample1} we determine the color of each bin by considering both shell stars and other halo stars that are inside the same spatial bin, in~the bottom panels, we show only those stars accreted from the shell-forming satellite (identified as described in Section \ref{subsec:historyCat}).
Thus, since in the bottom row we are isolating the stellar shells, it becomes more evident that, for~the galaxy in Figure~\ref{fig:metallicityShellsExample1}, stars in shells are more metal-rich than the overall stellar population of the host galaxy.

For the second study case, we consider a shell galaxy that exhibits $z=0$ shells composed of stars accreted from two different satellites.
The host galaxy in Figure~\ref{fig:metallicityShellsExample2} has a total stellar mass $\mathrm{M}_\mathrm{stars} = 1.3 \times 10^{12} \,\mathrm{M}_\odot$ and virial mass $\mathrm{M}_\mathrm{200crit} = 4.6 \times 10^{13}\, \mathrm{M}_\odot$. As before, we color each bin based on the median total metallicity of all star particles therein, and we find significantly more stellar shells at large radii, compared to the example in Figure~\ref{fig:metallicityShellsExample1}. However, similarly to the previous study case, stars in shells have higher total metallicities than the other halo stars at similarly large radii.
The two shell-forming satellites in Figure~\ref{fig:metallicityShellsExample2} were accreted between $\sim$6--7 Gyr ago, and they were stripped $\sim$3 and $\sim$2 Gyr ago, for satellites 1 and 2, respectively. Both of these merger events correspond to~relatively major mergers with $\mu_{\mathrm{stars}(t_{\mathrm{acc}})} = 0.15$ and $0.24$, and the satellites had radial infall trajectories, with~$v_r/|v| = -0.99$ and~$-0.84$ at $t_{\mathrm{acc}}$. The middle and bottom rows of Figure~\ref{fig:metallicityShellsExample2} depict only those stars accreted from each of these two satellites. On average, both satellites contribute stars with higher total metallicities than the average metallicity of stars in the host galaxy. Moreover, we find that the shells created by the first satellite are more metal-rich than those created by the second satellite.
This opens an interesting avenue for future observational studies---measurements of the metallicities of stars in shells could be used to identify shell galaxies with more than one shell-forming~progenitor.

In both Figures~\ref{fig:metallicityShellsExample1} and \ref{fig:metallicityShellsExample2}, we find a gradient in the metallicity of stars in shells, with outer shells having lower average Z than inner shells.
Studies of the satellites' infall trajectories show that shells in the current sample are often composed of stars stripped while the parent satellite passes close to the center of the host several times (see P17). Stars stripped during the satellite's first pericenter passage form the first generation of shells, with subsequent stripping events forming several generations of shells that are located at decreasing galactocentric distances compared to previous generations, (e.g.,~\cite{Salmonetal1990, Seguin&Dupraz1996, Bartoskovaetal2011, Cooperetal2011}). As a result, we expect the outer-most shells to be composed of stars that were initially located on the outskirts of the parent satellite, and, thus, they had lower average metallicities than stars closer to the satellite's core.
This explains the trends observed in Figures~\ref{fig:metallicityShellsExample1} and \ref{fig:metallicityShellsExample2}, with the first generation of shells having lower metallicities than shells located closer to the galaxy center.

\begin{figure}[H]
\centering
\includegraphics[width=1.0\textwidth]{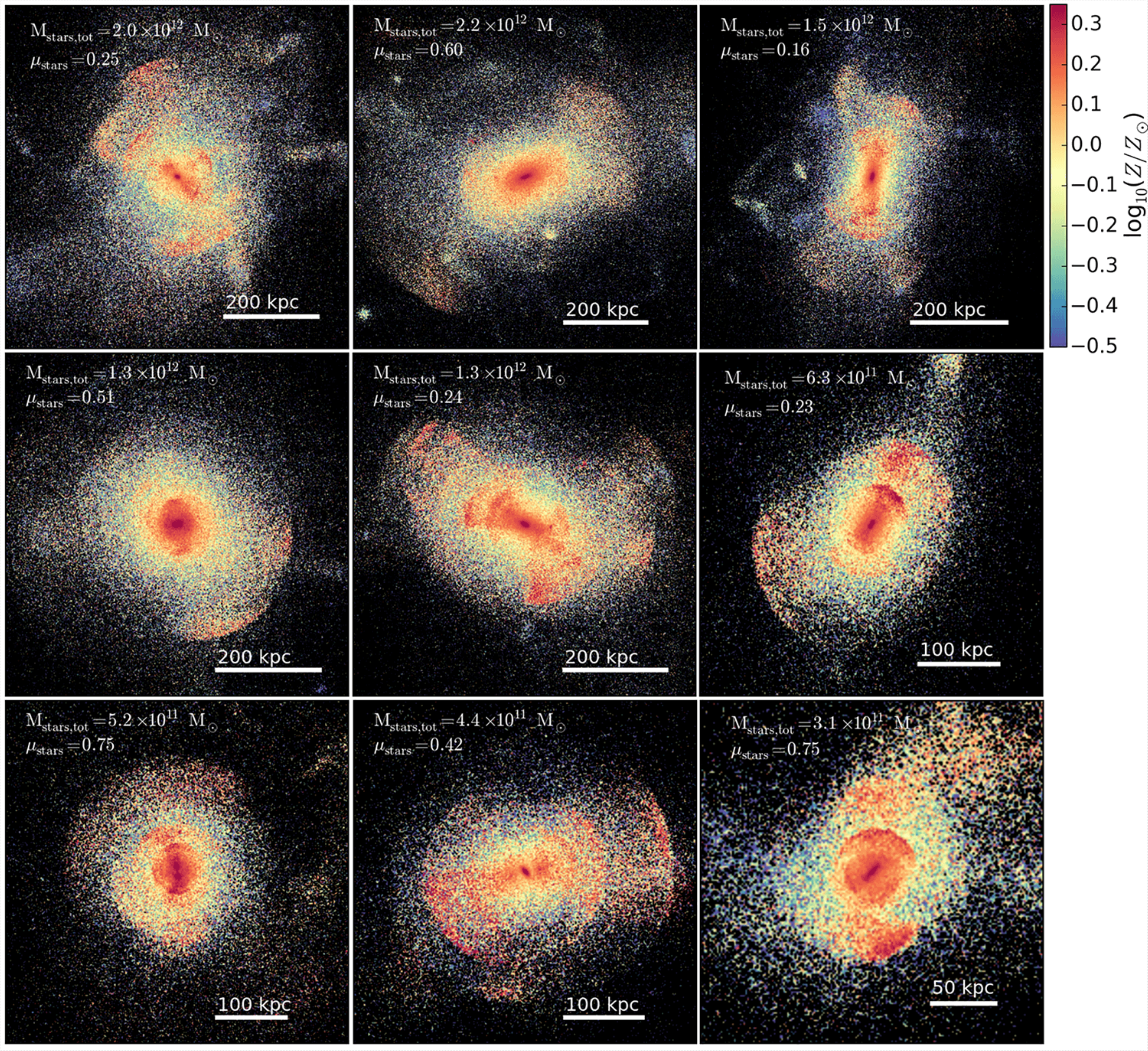}
\caption{Examples of nine different massive elliptical galaxies with shells from the Illustris simulation.
The color of each bin of area $\mathrm{A}_{\mathrm{bin}} = (1.5\, \mathrm{kpc})^2$ in the 2D histograms above corresponds to the median total metallicity (Z) of all star particles therein.
 The total stellar masses of these galaxies range from $\mathrm{M}_\mathrm{stars} = 3.1 \times 10^{11} \, \mathrm{M}_\odot$ to $\mathrm{M}_\mathrm{stars} = 2.2 \times 10^{12}\, \mathrm{M}_\odot$, while the virial masses are in the range $\mathrm{M}_\mathrm{200crit} \in  \left( 6.5 \times 10^{12} \, \mathrm{M}_\odot, \, 6.3 \times 10^{13} \, \mathrm{M}_\odot \right)$. For several of the shell galaxies in our sample, stellar shells are metal-rich compared to the total metallicity (Z) of stars in the outer stellar halo.}
\label{fig:metallicityShells9Panels}
\end{figure}

In Figure~\ref{fig:metallicityShells9Panels}, we show total metallicity Z-weighted stellar surface density maps for nine galaxies with shells in our sample. The virial masses of the host galaxies at $z=0$ span one order of magnitude: \mbox{$\mathrm{M}_\mathrm{200crit} \in \left( 6.5 \times 10^{12} \, \mathrm{M}_\odot, \;6.3 \times 10^{13} \, \mathrm{M}_\odot\right)$}, and total stellar masses are in the range \mbox{$\mathrm{M}_\mathrm{stars} \in \left( 3.1 \times 10^{11} \, \mathrm{M}_\odot,\; 2.2 \times 10^{12}\, \mathrm{M}_\odot\right)$}.
The examples presented in Figure~\ref{fig:metallicityShells9Panels} represent a selection of shell galaxies in Illustris that have outer shells
more metal-rich than the average metallicity of surrounding stars at similar galactocentric distances in the host galaxy. Nonetheless, we do not exclude that some galaxies could have shells composed of stars with similar or even lower metallicities than halo stars located at similar radii.

\begin{figure}[H]
\centering
\includegraphics[width=\textwidth]{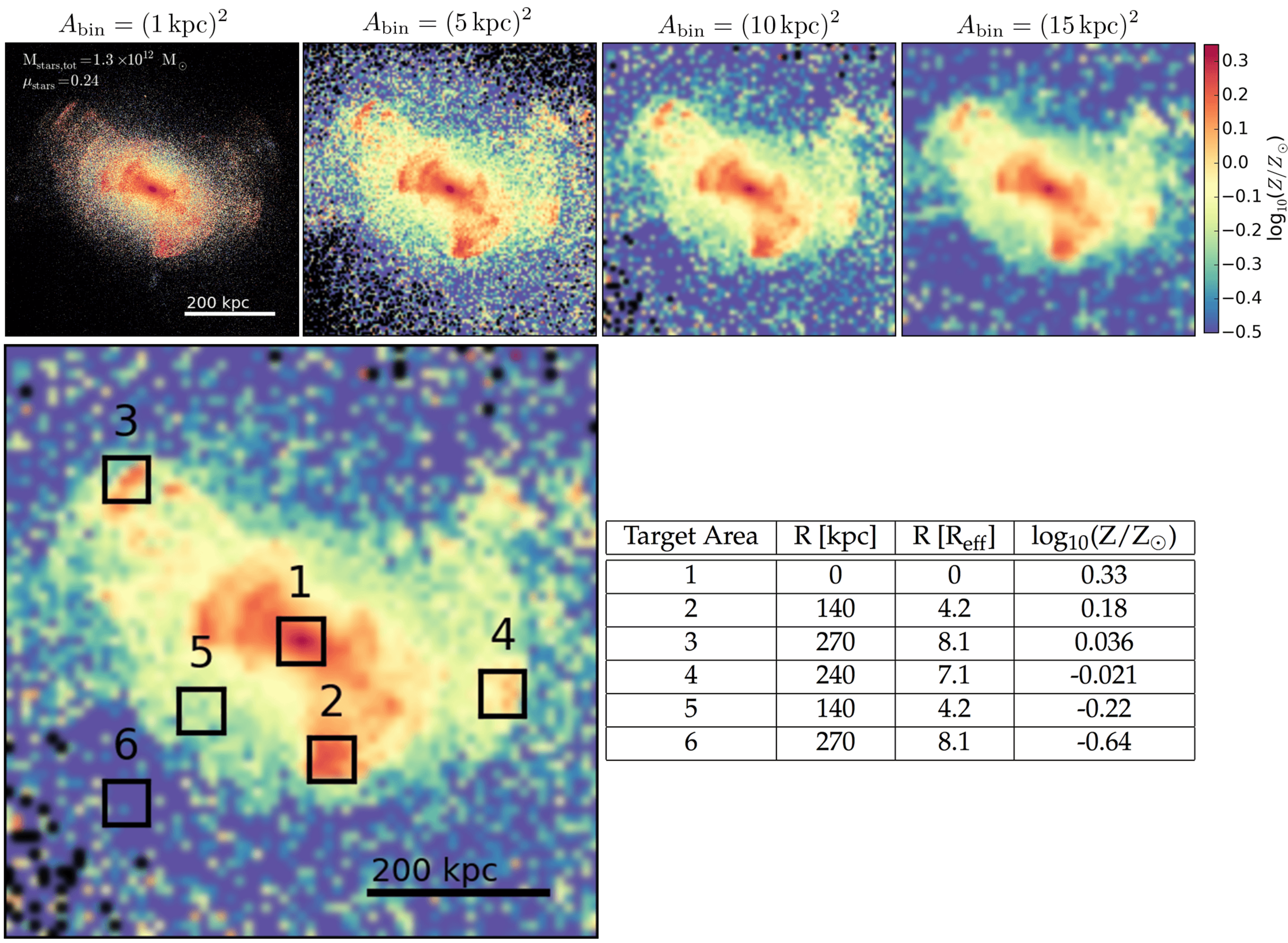}
\caption{\textls[-10]{\textbf{Top row:} Z-weighted stellar maps corresponding to the (\emph{y}--\emph{z}) projection
of all star particles inside the $z=0$ halo of a galaxy with shells with $\mathrm{M}_\mathrm{stars} = 1.3 \times 10^{12} \,\mathrm{M}_\odot$ and $\mathrm{R}_{\mathrm{eff}} = 33.5 \,\mathrm{kpc}$ (same~galaxy as in Figure~\ref{fig:metallicityShellsExample2}).
The resolution of the image stamps decreases from left to right, with the areas of 2D histogram bins given by: \mbox{$\mathrm{A}_{\mathrm{bin}} \in \left[ (1\, \mathrm{kpc})^2, \, (5\, \mathrm{kpc})^2, \, (10\, \mathrm{kpc})^2, \, (15\, \mathrm{kpc})^2    \right]$}. \textbf{Bottom row:} Zoom-in of the third image stamp in the top row, corresponding to a histogram with $\mathrm{A}_{\mathrm{bin}} = (10\, \mathrm{kpc})^2$. We mark six different target regions, each having a cross-section $\mathrm{A}_{\mathrm{target}} = (50\, \mathrm{kpc})^2$. \textbf{Table:} We provide the distance $R$ from the center of each of the target areas to the center of the galaxy in units of kpc and $\mathrm{R}_{\mathrm{eff}}$, respectively,~as~well as the median $\mathrm{log}_{10}(\mathrm{Z}/\mathrm{Z}_\odot)$ measured over all star particles inside each target area. Targets centered around outer shells (e.g., 2 and 3) have significantly higher median $\mathrm{log}_{10}(\mathrm{Z}/\mathrm{Z}_\odot)$ than targets at the same galactocentric distances but centered on regions without tidal features (e.g., targets 5 and 6).}}
\label{fig:metallicityShellsResolution}
\end{figure}

\textls[-10]{
In order to estimate the difference in total metallicity Z between outer shells and stars at similar galactocentric distances, and the resolution levels necessary to detect metal-rich shells such as those presented in Figure~\ref{fig:metallicityShells9Panels}, we next study a series of Z-weighted stellar maps at increasingly lower resolutions. In the top row of Figure~\ref{fig:metallicityShellsResolution}, we show the ($y$--$z$) projection of the same galaxy previously presented in Figure~\ref{fig:metallicityShellsExample2}, and
we vary the area of the 2D histogram bins from $\mathrm{A}_{\mathrm{bin}} = (1\, \mathrm{kpc})^2$ to $\mathrm{A}_{\mathrm{bin}} = (5\, \mathrm{kpc})^2$, $(10\, \mathrm{kpc})^2$, and finally, $(15\, \mathrm{kpc})^2$. This galaxy has an effective radius $\mathrm{R}_{\mathrm{eff}} = 33.5 \,\mathrm{kpc}$, so the right-most stellar map in Figure~\ref{fig:metallicityShellsResolution} corresponds to a low-resolution image, with the length of each bin being \linebreak $\mathrm{L}_{\mathrm{bin}} \approx \mathrm{R}_{\mathrm{eff}}/2\,$.
Some of the fine structure associated with tidal debris is completely erased for \linebreak $\mathrm{A}_{\mathrm{bin}} > (5\, \mathrm{kpc})^2$, and the number of detectable shells drops significantly as we lower the resolution.
Nonetheless, there is sufficiently high contrast between the metallicities of some of the outer shells and the metallicities of stars located at similar radii in the halo, such that several outer shells can
still be observed even for $\mathrm{A}_{\mathrm{bin}} \simeq (15\, \mathrm{kpc})^2$.
In the bottom row of Figure~\ref{fig:metallicityShellsResolution}, we~present a larger version of the image stamp with $\mathrm{A}_{\mathrm{bin}} = (10\, \mathrm{kpc})^2$, on which we mark six different targets, each having an area $\mathrm{A}_{\mathrm{target}} = (50\, \mathrm{kpc})^2$. In the table accompanying the figure, we provide galactocentric distances for the centers of each of the~six targets in units of kpc and $\mathrm{R}_{\mathrm{eff}}$, as well as the median $\mathrm{log}_{10}(\mathrm{Z}/\mathrm{Z}_\odot)$ computed for all star particles located inside each target area\footnote{We assign stars to each target area by selecting all star particles gravitationally bound to the galaxy and located inside a rectangular box with ($y$--$z$) square cross-section of area $\mathrm{A}_{\mathrm{target}} = (50\, \mathrm{kpc})^2$ and infinite length.}.
The~median $\mathrm{log}_{10}(\mathrm{Z}/\mathrm{Z}_\odot)$ is highest for target area~1, which is centered on the core of the galaxy. Areas $2$--$4$ are targeting three different shells located at galactocentric distances: $\mathrm{R}_2 = 4.2 \, \mathrm{R}_{\mathrm{eff}}$, $\mathrm{R}_3 = 8.1 \, \mathrm{R}_{\mathrm{eff}}$, and $\mathrm{R}_4 = 7.1 \, \mathrm{R}_{\mathrm{eff}}$, respectively. The~median $\mathrm{log}_{10}(\mathrm{Z}/\mathrm{Z}_\odot)$ of all stars in these target areas varies between $\mathrm{log}_{10}(\mathrm{Z}/\mathrm{Z}_\odot) = 0.18$ to $-0.021$.
Despite the median total metallicities of stars in outer shells being lower than the median Z of stars near the galaxy center (where $\mathrm{log}_{10}(\mathrm{Z}/\mathrm{Z}_\odot) = 0.33$), there is still a significant contrast in metallicity between targets at similar radial distances depending on whether they are centered on a shell or not. For~example, for~target~2, we measure a median $\mathrm{log}_{10}(\mathrm{Z}/\mathrm{Z}_\odot) = 0.18$, while target 5 is positioned at the same galactocentric distance ($\mathrm{R} = 4.2 \, \mathrm{R}_{\mathrm{eff}}$), in a region of the stellar halo where we detect no tidal features, and it has a median $\mathrm{log}_{10}(\mathrm{Z}/\mathrm{Z}_\odot) = -0.22$.
This corresponds to a metallicity contrast $\Delta \mathrm{log}_{10}(\mathrm{Z}/\mathrm{Z}_\odot) = 0.4$ between targets 2 and 5.
We can similarly compare targets~3 and 6, with the former being centered on a shell in the upper left quadrant and having median $\mathrm{log}_{10}(\mathrm{Z}/\mathrm{Z}_\odot) = 0.036$, while~target 6 is positioned at a similar radial distance ($\mathrm{R} = 8.1 \, \mathrm{R}_{\mathrm{eff}}$)  but in a low stellar density region of the halo, where the median $\mathrm{log}_{10}(\mathrm{Z}/\mathrm{Z}_\odot) = -0.64$, leading to a contrast $\Delta \mathrm{log}_{10}(\mathrm{Z}/\mathrm{Z}_\odot) \approx 0.7$ between targets 3 and 6.  In~addition, we find that the median total metallicities of outer shells ($\mathrm{log}_{10}(\mathrm{Z}/\mathrm{Z}_\odot) = 0.036$ and $-0.021$ for targets~3 and 4, respectively) are lower than the median metallicity for the shell located at a smaller radius, near~target number 2 ($\mathrm{log}_{10}(\mathrm{Z}/\mathrm{Z}_\odot)=0.18$). These~results are in agreement with the metallicity gradients in shells observed in Figures~\ref{fig:metallicityShellsExample1} and~\ref{fig:metallicityShellsExample2}: outer~shells are less metal-rich than inner shells. Due to the extremely low surface brightness of shells, reaching the precision levels above is challenging for current observations, though~not technically~impossible.}

\begin{figure}[H]
\centering
\includegraphics[width=\textwidth]{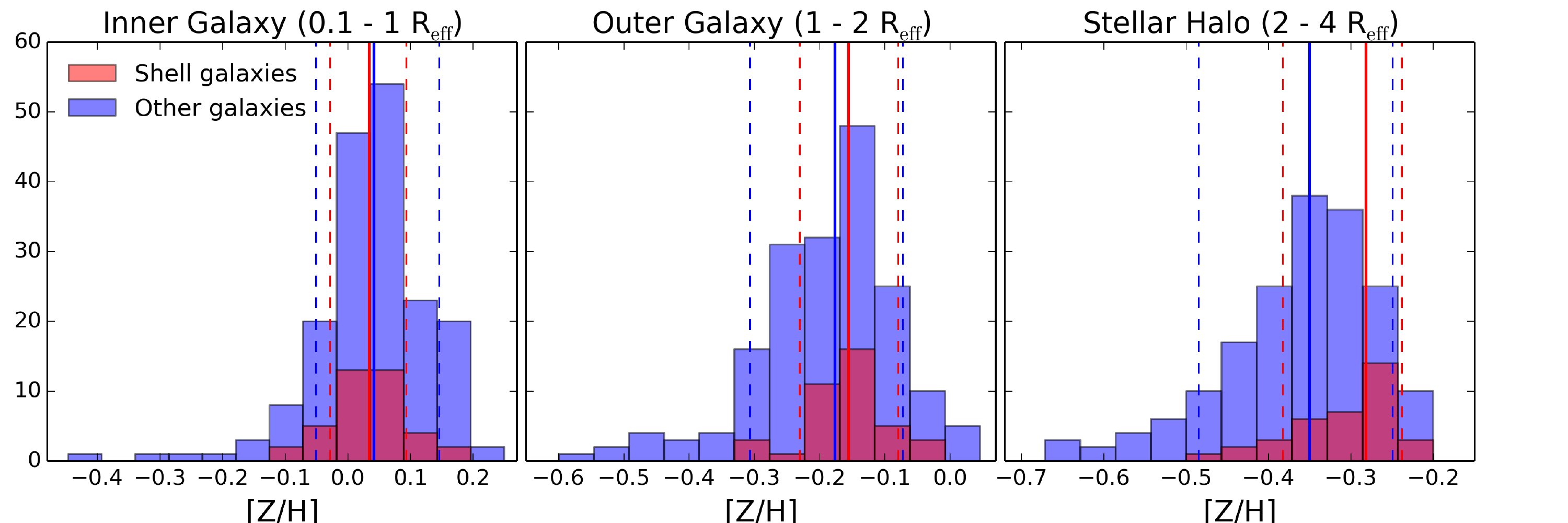}
\caption{Histograms of the average logarithmic metallicities of stars in a sample of 220 massive elliptical galaxies. The shell galaxies identified in P17 are shown in red, with the other galaxies in the sample shown in blue. We measure [Z/H] within three radial ranges, based on the effective radius of each galaxy, $\mathrm{R}_{\mathrm{eff}}$, defined as the radius that contains half of the total light.
The median [Z/H] for all shell/non-shell galaxies is marked by a continuous line, while dashed lines mark the $10\%$ and~$90\%$ percentiles.
The average total stellar metallicities are similar for shell and non-shell galaxies at galactocentric distances
$< 2\, \mathrm{R}_{\mathrm{eff}}$. In the extended stellar halo, we find that shell galaxies in our sample have higher average metallicities (shallower metallicity gradients) than galaxies without shells.
}
\label{fig:histograms}
\end{figure}

Next, we quantify the difference in the median stellar metallicities of shell vs. non-shell galaxies by measuring the mass-weighted logarithmic metallicities [Z/H] of stars around all 220 galaxies in our sample (see Figure~\ref{fig:histograms}).
Following the definitions adopted in \cite{Cooketal2016}, we measure [Z/H] within three radial ranges: the inner galaxy (\mbox{$0.1$--$1 \; \mathrm{R}_{\mathrm{eff}}$}), outer galaxy (\mbox{$1$--$2 \; \mathrm{R}_{\mathrm{eff}}$}), and stellar halo (\mbox{$2$--$4 \; \mathrm{R}_{\mathrm{eff}}$}). The~average effective radius of the entire sample is \mbox{$20.9\, \mathrm{kpc}$}, with shell galaxies having a higher average (\mbox{$\mathrm{\bar{R}}_{\mathrm{eff}} = 24.5\, \mathrm{kpc}$}) than galaxies without shells (\mbox{$\mathrm{\bar{R}}_{\mathrm{eff}} =  20.1\, \mathrm{kpc}$}). We do not find a significant difference in the average metallicities of stars within \mbox{$\sim$2  $\mathrm{R}_{\mathrm{eff}}$} of the galaxies' centers. However, in the outer regions of the stellar halo (\mbox{$2$--$4 \; \mathrm{R}_{\mathrm{eff}}$}), shell galaxies have higher average metallicities than galaxies without shells: the median values for the two distributions are \mbox{$[Z/H]_{\mathrm{shells}}=-0.28$} and \mbox{$[Z/H]_{\mathrm{non\text{-}shells}}=-0.36\,$}. In~turn, if stars on the outskirts of shell galaxies are more metal-rich, this implies that the metallicitiy gradients of galaxies with shells could be shallower than those of non-shell galaxies. Despite the higher average [Z/H] for shell galaxies, we find that there is significant scatter in the metallicity distributions for galaxies in our sample.
Most galaxies do not show azimuthal symmetry, and, therefore, by computing [Z/H] averaged over radial profiles, we are ignoring the relative influence of different stellar substructures present at similar radii in the halo.
We thus expect measurements of stellar metallicities targeted towards regions with significant stellar overdensities or clumpy substructures (see, for example, Figure~\ref{fig:metallicityShellsResolution}) to be better suited at identifying new shells and characterizing the mergers producing them than measurements of overall metallicity gradients or [Z/H] averaged over large radial bins.

In P17, we show that most merger events responsible for forming $z=0$ shells in massive elliptical Illustris galaxies can be described by an order-zero recipe, involving only three parameters: stellar mass ratio ($\mu_{\mathrm{stars}}$), radial velocity ratio ($v_r/|v|$), and accretion time ($t_{\mathrm{acc}}$).
Satellites that successfully form $z=0$ shells correspond to mergers with stellar mass ratios $\mu_{\mathrm{stars}} \gtrsim 1:10$
and are accreted on very low angular momentum orbits, about $\sim$4--8 Gyr ago.
While the mean total mass ratio ($\mu_{\mathrm{total}} = \mathrm{M}_{\mathrm{satellite}}^{\mathrm{total}} / \mathrm{M}_{\mathrm{host}}^{\mathrm{total}}$) of shell-forming merger events in P17 is lower than the mean stellar mass ratio $(\mu_{\mathrm{stars}})$, both $\mu_{\mathrm{total}}$ and  $\mu_{\mathrm{stars}}$ are biased towards close-to-major mergers. Due to dynamical friction being more efficient at radializing the orbits of satellites involved in close to $1:1$ mergers (see, e.g., \cite{BoylanKolchinetal2013, Amorisco2017}),
massive satellites are allowed to probe a wider range of impact parameters and still be successful at forming shells (see discussion in P17). On the other hand, in order to produce shells in massive ellipticals, small satellites need very fine-tuned, almost perfectly radial orbits. Thus, P17 find that shells in massive galaxies are produced more frequently by major or close-to-major mergers than by minor mergers.

Previous studies have shown that major mergers can deposit material closer to the center of the host, unlike minor mergers which mostly contribute to the ex-situ stellar fraction at large galactocentric radii \cite{Rodriguez-Gomezetal2016,Amorisco2017}. Major mergers with shell-forming satellites containing metal-rich stars will thus deposit high-metallicity material over a wide range of radii---from the inner regions of the halo out to distances~$\gtrsim 100$ kpc.
As a result, there are two competing effects that need to be considered when estimating the metallicity of outer shells relative to the metallicity of halo stars at similar radii.
On~one hand, the outer-most shells (or the first generation of shells) are composed of stars stripped from the outskirts of the satellite's stellar halo during its first pericenter approach, while subsequent shells at lower radii correspond to more metal-rich stars that were located closer to the center of the shell-forming progenitor. This has the effect of generating a metallicity gradient in the shells, with outer shells being more metal-poor than inner shells (as observed in Figures~\ref{fig:metallicityShellsExample1} and \ref{fig:metallicityShellsExample2}).
On the other hand, shells extend to extremely large distances in the low surface brightness and metal-poor outskirts of the host's stellar halo. Despite the fact that the very first $\sim$1--2 shells at large radii are generally not very metal-rich, the next shells will start to contain relatively high-Z stars stripped from closer to the center of the satellite. After the merger event, these stars are located in shells extending to much larger effective radii in the host galaxy compared to their initial galactocentric distances with respect to the center of the shell-forming progenitor.
Therefore, in the case of major mergers, many of the outer shells can be on average more metal-rich than halo stars located at similar radii in the host galaxy, and~we observe this effect in several of the shell galaxies in our sample (see Figure~\ref{fig:metallicityShells9Panels}).

When averaging stellar metallicities over large radial bins, we find that shell galaxies have higher median [Z/H] than non-shell galaxies in the outer regions of the stellar halo ($2$--$4\, \mathrm{R}_{\mathrm{eff}}$).
Hirschmann~et~al.~\cite{Hirschmann2015} used zoom hydrodynamical simulations of 10 galaxies with masses \linebreak above~$6\times10^{12}\,~\mathrm{M}_\odot$, probing a similar mass range as our current study. They found that systems that have undergone major mergers have significantly shallower metallicity gradients at galactocentric distances $>2\,\mathrm{R}_{\mathrm{eff}}$ compared to systems dominated by minor mergers.~For Illustris galaxies, Cook~et~al.~\cite{Cooketal2016} showed that galaxies with high ex-situ fractions tend to have less steep metallicity gradients (in the radial range $ 2$--$4 \, \mathrm{R}_{\mathrm{eff}}$) than those with smaller accreted fractions, yet the metallicity gradients are not strongly correlated with the mean merger mass-ratio.~As a result, it can be difficult to draw conclusions based solely on the distribution of overall metallicity gradients in shell galaxies, and~precision metallicity measurements targeted around individual shells can provide more information about the shell-forming progenitor.

Based on the results presented above, we expect even outer shells (formed from material stripped from the outskirts of satellites) to be relatively metal-rich if the shells are produced by a massive parent satellite. Therefore, observations of metal-rich shells could indicate that a high fraction of the shells observed in massive early-type galaxies have a major merger origin. In this way, precision metallicity measurements targeting stars in and around the shells can provide additional information about the formation processes of shell galaxies.

\section{Summary}

{In the current study}, we investigate the average total metallicity of stellar shells around massive elliptical galaxies from the Illustris simulation.
We present several shell galaxies for which we identify shells in both configuration space ($x$--$y$, $y$--$z$, $z$--$x$) and phase space ($r$--$v_r$).
Stars in the outer shells of these galaxies have higher total metallicities than other stars at similar radii in the host galaxies. This result suggests that the shells studied in this paper are the by-product of major mergers.
Moreover, in the particular case of a galaxy with two different shell-forming progenitors, we find a significant difference in the metallicity of the shells produced by each of the two satellites.
This~indicates that high-precision metallicity measurements could potentially identify shell galaxies with multiple progenitors based on their different metallicity signatures.

Previous analytical and numerical studies (e.g., \cite{Dupraz&Combes1987, Cooperetal2011})
have indicated that several generations of shells are necessary to explain the wide range of shell radii measured in observations of early-type galaxies \cite{Wilkinsonetal1987}. In agreement with these results, we find that shells in Illustris are often formed by stripping stars while the parent satellite performs several pericenter passages.
Stars in the first generation of shells (the outer-most shells) correspond to the least bound stars on the outskirts of the progenitor galaxy and, thus, they are the first stars stripped from the satellite.
We find a gradient in the average metallicity of shells, with outer shells in our two study cases having lower metallicities than inner shells.
Since stars on the outskirts of galaxies tend to have lower metallicities, first generation shells would be expected to be more metal-poor than subsequent generations, in agreement with our~findings.

Finally, we investigate the average mass-weighted metallicities [Z/H] in three different radial bins corresponding to the inner galaxy, outer galaxy and stellar halo.
Both galaxies with and without shells in our sample seem to have a similar average [Z/H] within 2 $\mathrm{R}_{\mathrm{eff}}$ of the galaxy center. We find, however, a bias towards a higher average [Z/H] for stars in the outer stellar halos of shell galaxies.

The results presented in this paper indicate that several of the shell galaxies in Illustris have outer shells more metal-rich than the average metallicity in the stellar halo at similar radii. We predict that multiple shell-forming progenitors, as well as different generations of shells, could leave specific signatures in the metallicity distribution of stars in shells.
In addition to having the potential to find previously undetected shells, future metallicity observations could unveil information about the formation processes of shell galaxies, such as distinguishing between minor/major mergers producing shells or the number of different satellites responsible for the observed structure.

\vspace{6pt}


\acknowledgments{A.R.P. wishes to thank the referees for helpful, constructive comments that improved this paper.
A.R.P. would like to thank Ben Cook for useful discussions on stellar metallicities in Illustris, as well as Paul Torrey, Charlie Conroy, and Jieun Choi for answering questions related to metallicities. A.R.P. also thanks Vicente Rodriguez-Gomez for the merger trees in Illustris, and Jane Huang, John Lewis and Luke Kelley for useful comments on the draft. Special thanks go to the organizers of the conference ``On the Origin (and Evolution) of Baryonic Halos''.}

\authorcontributions{A.R.P., A.P., N.C.A., and L.H. designed the research; A.R.P, A.P., and N.C.A. visually identified shell galaxies using stellar surface density maps; A.R.P. developed stellar history catalogs and post-processing tools, identified shell-forming progenitors and performed the numerical analysis; A.R.P. wrote the paper.}


\conflictsofinterest{The authors declare no conflict of interest.}
\bibliographystyle{mdpi}




\end{document}